\documentstyle[preprint,epsfig,aps]{revtex}
\newcommand{\bea}{\begin{eqnarray}}
\newcommand{\eea}{\end{eqnarray}}
\newcommand{\be}{\begin{equation}}
\newcommand{\ee}{\end{equation}}

\begin{document}
\draft
\preprint{\begin{tabular}{l}
\hbox to\hsize{July, 1998 \hfill KAIST-TH 10/98}\\
\hbox to\hsize{hep-ph/9811211 \hfill SNUTP 98-007
}\\[5mm] \end{tabular} }

\bigskip

\title{ Complete analysis of photino-mediated lepton flavor \\ violations  
in generalized supersymmetric models}
\author{Yeong Gyun Kim, ~~~~~Pyungwon Ko  \\
 Jae Sik Lee, ~~~and ~~~ Kang Young Lee }
\address{Department of Physics\\
Korea Advanced Institute of Science and Technology 
\\ Taejon 305-701, Korea}
\maketitle
\begin{abstract}
We consider lepton flavor violations (LFV) mediated by photino as a result of
the nondiagonal slepton mass matrices in general supersymmetric models.  
Using the experimental upper bounds on $l \rightarrow l^{'} + \gamma$ and
$\mu^- + {\rm Ti} \rightarrow e^- + {\rm Ti}$ 
as constraints on the flavor changing slepton mass
insertions, we predict the possible ranges of the upper limits on
the branching ratios of other LFV processes such as 
$l \rightarrow 3l',
~\tau \rightarrow l_{i\neq 3} + \pi^0 ({\rm or}~\eta, \rho^0, \phi), 
~Z^0 \rightarrow l_i \overline{l}_{j\neq i}$,
and muonium $\rightarrow$ antimuonium conversion.                     
Most of these decays are expected to occur with small branching ratios
far below  the current or future experimental search limits.
We also derive constraints on the 
flavor conserving mass insertions from the anomalous magnetic moments of 
leptons.
\end{abstract}


\newpage
\narrowtext
\tighten

\section{Introduction}

At present, the minimal supersymmetric standard model (MSSM) is widely 
considered as the leading candidate for the physics beyond the standard model 
(SM) \cite{intro}. 
It can solve the gauge hierarchy problem by supersymmetrizing
the SM (with an addtional Higgs doublet).  
The boson loop contribution to Higgs mass is cancelled by the
fermion loop contribution, the latter of which comes in with opposite sign to
the former.  In doing so, the particle spectrum of the theory becomes doubled
compared to that of the SM. One expects a lot of new scalar particles 
(superpartners of the SM fermions) and new fermions (superpartners of the
SM gauge bosons and Higgs).  These new particles should have masses around 
$O(100)$ GeV $- O(1)$ TeV in order to solve the gauge hierarchy problem 
in terms of softly broken supersymmetry.  Nice features of the supersymmetric
theories are their calculability using perturbation theory, and the decoupling
nature of the loop effects of new (super)particles on various electroweak 
observables, except for the dangerous supersymmetric (SUSY) flavor changing 
neutral current (FCNC) and SUSY CP problems (which will be dicussed shortly 
in more detail).  Therefore the successful predictions of the SM do not 
change very much even if we have doubled spectrum of particles in SUSY 
theories modulo SUSY FCNC and CP problems.       

However, in generic supersymmetric (SUSY) models, one has to pay for this 
extra symmetry. First of all, the lepton family numbers ($L_{i=e,\mu,\tau}$) 
and the baryon number ($B$) are no longer conserved as in SM. 
One can write down renormalizable superpotential which violates the $L_i$ 
and $B$ numbers and leads to too fast  proton decay in conflict with the
observation.  Secondly, the soft mass terms for sfermions can lead to 
large FCNC unless certain conditions are met.  
In most phenomenological SUSY models, one solves the first problem by 
assuming $R-$parity conservation by hand. The second problem  (SUSY FCNC) 
is solved by assuming that either (i) the sfermion mass matrices are 
proportional to the unit matrix in the flavor space \cite{hall86}
\cite{masiero}, (ii) the sfermion mass matrices are proportional to the 
corresponding fermion mass matrices so that both can be diagonalized 
simultaneously \cite{nir}, or (iii) assuming that the first two generation
sfermions are highly degenerate and very heavy ($gtrsim 50$ TeV 
$>> M_{SUSY}$ so that they basically decouple) \cite{nelson}. 
In the minimal SUGRA models with the flat K\"{a}hler metric at 
$M_{\rm Planck}$ scale, the first condition can be met, namely the squarks, 
sleptons and  Higgs are all degenerate at the Planck scale.  
However, when one evolves the sfermion mass parameters to the electroweak 
scale using renormalization group (RG), the off-diagonal elements of the 
squark mass matrices are  induced in a calculable manner, although there is 
no LFV induced at low energy. 
Moreover, the condition of the flat K\"{a}hler metric is a strong assumption 
which may not be true in general. For example, SUGRA radiative corrections
to the boundary conditions at $M_{\rm Planck}$ scale induce generically 
$O(\sim 10\%)$ off-diagonal sfermion mass matrix elements \cite{kchoi}. 
Therefore, one can imagine certain amount of nondiagonal sfermion mass matrix 
elements at the electroweak scale in general. In the lepton sector of the 
minimal SUGRA model, there is no LFV as in the SM since neutrinos are 
massless. As an example, consider the minimal SUGRA model with the flat 
K\"{a}hler metric. Then the scalar masses are universal (being $m_0^2$) at 
the Planck scale, whereas at the weak scale the sfermion masses change as  
\begin{equation}
( m_{\tilde{d}}^2 )_{LL} (\mu = m_Z ) = m_d m_d^{\dagger} + m_0^2 + 
c_d m_{u} m_{u}^{\dagger}, 
\end{equation}
as a result of renormalization \cite{hall86}. Because of the last term 
containing $m_u m_u^{\dagger}$, it is not possible to diagonalize $m_d 
m_d^{\dagger}$ and $( m_{\tilde{d}}^2 )_{LL} $ simultaneously. This leads 
to the flavor changing gluino-quark-squark vertices, which can contribute 
to various low energy  FCNC processes. Also this $(LL)$ mixing induces the
$LR$ and $RR$ mixings in the minimal SUGRA models. For sleptons, on the 
other hand,  we have
\begin{equation}
( m_{\tilde{l}}^2 )_{LL} (\mu = m_Z ) = m_l m_l^{\dagger} + m_0^2 + 
c_l m_{\nu} m_{\nu}^{\dagger}, 
\end{equation}  
and massless neutrinos (namely, absence of righthanded neutrinos) 
imply that there is no generation mixing in the slepton mass matrix.  
Since $ LR$ and $RR$ transitions are proportinal to the $(LL)$ mixing, 
there will be no lepton family number violation in the
minimal SUGRA models. However, if there is LFV at high energy scale 
(e.g., in Supersymmetric Grand Unification Theories (SUSY GUT)) 
\cite{susygut} or if right-handed neutrinos are included in the SUGRA 
\cite{hisano}, there can be generic LFV at electroweak scale. 

In view of this, it is important to see how large deviation from the above
conditions (i) and (ii) are allowed in the general SUSY models by the various 
FCNC processes at low energy. Such studies have been done previously by 
several authors already, mainly on the gluino--mediated FCNC in the quark 
sector and the photino--mediated $l_i \rightarrow l_{j\neq i} + \gamma$ 
\cite{masiero}. Basically deviations between the first and the second families
should be very small. In terms of a dimensionless parameter defined as 
\begin{equation}
( \delta^{l} )_{AB} \equiv \Delta^l )_{AB} / m_{\tilde{l}}^2,
\end{equation}
 where $ m_{\tilde{l}}^2$ is a suitable average of the slepton masses, 
the condition that deviations between the first and the second families
should be very small can be represented as following constraints 
\cite{masiero}:   
\begin{equation}
(\delta_{12}^{l} )_{LL} = O(10^{-3}), 
~~~{\rm and}~~~
(\delta_{12}^{l} )_{LR} = O(10^{-6}).
\end{equation}
On the other hand, the deviations involving the third family are more 
loosely constrained :
\begin{equation}
(\delta_{13(23)}^{l} )_{LL} = O(1-10), 
~~~{\rm and}~~~
(\delta_{13(23)}^{l} )_{LR} = O(10^{-2}).
\end{equation}
This is in part due to the less precise experimental informations on 
various FCNC processes involving the third family. But there are many 
interesting possibilities for which one can treat the third famly in a 
different manner from the first two families. In such theories, one may 
expect larger deviations from the degeneracy in general, and thus expect FCNC 
processes with branching ratios that may be accessible in the near future.  

In this work, we mainly concentrate on the photino-mediated FCNC processes
in the lepton sector, which are almost parallel to the work by  Gabbiani 
{\it et al.} \cite{masiero}. Namely, we assume that the slepton mass matrices 
are not diagonal in the basis where $\tilde{l}_i -l_j -\tilde{\gamma}$ 
vertex is flavor diagonal. In order to simplify the analysis, we make an 
assumption that the lightest superparticle (LSP) is a photino 
($\tilde{\gamma}$), and other neutralinos are fairly massive so that their 
effects are negligible compared to the LSP effects considered in this 
work.  Finally, we assume that the off--diagonal mass matrix elements of 
sleptons are  small enough that the mass insertion approximations are 
applicable. All of these assumptions are the same as Ref.~\cite{masiero}, 
except for the photino mediated LFV instead of gluino mediated FCNC.  
In the case of glino-mediated FCNC, the neutralino effects will be 
generically suppressed by $\alpha_2/\alpha_s$, so that one can safely 
ignore the neutrlino-mediated FCNC. For the case of LFV, all the couplings 
of four neutralinos will be the same order of magnitude, and all the 
neutralino contributions to LFV should be included at the same time
in principle. However, we assume that the photino is the LSP and other 
neutralinos are heavy and can be ignored in order to simplify our analysis.
It would be straighforward, although tedious, to include 4 neutralinos
altogether and make more complete our analysis.

There are a few differences between Ref.~\cite{masiero} and our 
work. First of all, we can restrict the allowed regions of the FC mass 
insertion by considering different processes. Different processes provide
independent constraints from each other, and we need not make an assumption
that there is no fortuitous cancellation between $\delta_{LL}$ and 
$\delta_{LR}$, and so on. In the limit the  light photino dominates the 
LFV, we can even predict the upper bound on some LFV decays in a completely
model independent fashion. It is straightforward to relax this assumption and 
include all the four neutralino contributions to LFV, if necessary.  
We consider all the LFV processes 
that are studied experimentally at present. 
We consider LFV decays of $Z^0$ 
gauge boson, and processes involving two leptons and two quarks, such as 
$\mu^- + Ti \rightarrow e^- + Ti$, 
and $\tau \rightarrow \mu (e) + $(a neutral meson)
as well as processes involving four leptons and the LFV radiative decays. 

Secondly, the authors of Ref.~\cite{masiero} derived constraints 
on the flavor conserving mass insertion 
$\delta_{ii}^l$ from the requirement that the SUSY 
one-loop contribution to the  lepton mass (one loop diagram with an 
insertion of $\delta_{ii}^l$) is smaller than the actual lepton mass 
($\Delta m_l^{\rm SUSY} < m_l^{\rm exp}$). However, we regard this condition 
as an improper one, since the particle mass cannot be predicted by SM or 
SUSY models. On the contrary, it turns out that 
the anomalous magnetic moment of a lepton ($a_{l} 
\equiv (g - 2)/2$) can provide more meaningful and 
stronger bounds on $\delta_{ii}^l$. 

This paper is organized as follows. 
In Sec.~II, we construct the effective lagrangian for $\Delta L_i = 1$ and 2. 
The results form the basis for the calculations of transition rates for 
various LFV processes in the Sec.~III. 
Constraints on the flavor conserving mass insertions from the anomalous 
magnetic moment  are derived in Sec.~IV, and the results are summarized 
in Sec.~V. 
 
\section{Effective Lagrangian for $\Delta L_i$ = 1 and 2 }

\subsection{${\cal L}_{\rm eff} (4 l)$ for $\Delta L_i = 1$}

Let us first derive the effective Lagrangian for $\Delta L_i = 1$. 
A complete basis for $\Delta L_i = 1$ effective Lagrangian is

\begin{equation}
{\cal L}_{\rm eff}^{\Delta L_i = 1} (4 l) = \sum_{i=3,5,7} 
\left[ ~C_i O_i + C_i^{'} O_i^{'} ~\right], 
\end{equation}

\noindent
where

\begin{eqnarray}
O_3 & = & \overline{l_j} \gamma_{\mu} P_L l_{i}~\sum_{k} ~
\overline{l_k} \gamma^{\mu} P_L l_k,
\nonumber  \\
O_5 & = & \overline{l_j} \gamma_{\mu} P_L l_{i}~\sum_{k}~
\overline{l_k} \gamma^{\mu} P_R l_k,
\nonumber   \\
O_7 & = & {e \over 8 \pi^2} ~m_{i} \overline{l_j} \sigma^{\mu\nu} P_R l_{i}
F_{\mu\nu}.
\end{eqnarray} 

\noindent
The operators $O_i^{'}$'s and the associated Wilson coefficients 
$C_i^{'}$'s are obtained from $O_i$'s and $C_i$'s by the exchange $L 
\leftrightarrow R$. Evaluating the Feynman diagrams in Figs.~
1 (the box diagrams) and 2 (the penguin diagrams), and matching the full 
amplitudes with those in the effective theory, we get 

\begin{eqnarray}
C_3 & = &  {2 \alpha^2 \over m_{\tilde{l}}^2} \left( \delta_{ji}^l 
\right)_{LL} \left [ P_1 (x) - 4  B_1 (x) - 2 B_2 (x)  \right] ,
\nonumber  \\
C_5 & = &  {2 \alpha^2 \over m_{\tilde{l}}^2} \left( \delta_{ji}^l 
\right)_{LL} \left[ P_1 (x) + 4 B_1 (x) + 2 B_2 (x) \right],
\nonumber   \\
C_7 & = & {2 \alpha \pi \over m_{\tilde{l}}^2}~\left[ \left( \delta_{ji}^l 
\right)_{LL} M_3 (x) + \left( \delta_{ji}^l \right)_{LR} {m_{\tilde{\gamma}}
\over m_{i}} M_1 (x) \right]. 
\end{eqnarray} 

\noindent
We have neglected the final lepton mass $m_j$ in the above expression, and
$x \equiv m_{\tilde{\gamma}}^2 / m_{\tilde{l}}^2$~\footnote{Strictly speaking,
the  photino LSP implies that  $x<1$. However, we consider the case $x > 1$ 
as well, since it would give a rough estimate of neutralino-mediated LFV's
in case  the photino is no longer the LSP.}. 
As noted in Ref.~\cite{masiero}, the $Z-$penguin contributions to 
$\mu \rightarrow 3 e$ {\it etc.} are suppressed compared to the above 
by a factor of $m_l^2 / M_Z^2$, and thus were safely ignored. 
Note that the $\delta_{LR}$ and $\delta_{RL}$ contribute only to $O_7$,
and not to $O_{3,5}$, when we keep only dimension$-$6 operators in our 
effective  theory.  

The functions $B_i$'s (from the box diagrams, Fig.~1), $P_i$'s (from the 
penguin diagrams, Fig.~2) are defined in Ref.~\cite{masiero}, and shown 
below for completeness :

\begin{eqnarray}
B_1 (x) & = & {1 + 4x - 5x^2 + 4 x \ln (x) + 2 x^2 \ln (x) \over 
8 (1 -x)^4} ,
\nonumber   \\
B_2 (x) & = & x { 5-4x - x^2 + 2 \ln (x) + 4 x \ln (x) \over
2 (1-x)^4} , 
\nonumber    \\
P_1 (x) & = & {1 - 6x + 18 x^2 - 10 x^3 - 3 x^4 + 12 x^3 \ln (x) \over
18 (x-1)^5 } ,
\nonumber   \\
P_2 (x) & = & {7 - 18 x + 9 x^2 + 2 x^3 + 3 \ln (x) - 9 x^2 \ln (x) \over
9 (x-1)^5} ,  
 \nonumber    \\
M_1 (x) & = & 4 B_1 (x) ,
 \nonumber   \\
M_3 (x) & = & {-1+9x +9x^2 - 17 x^3 + 18 x^2 \ln (x) + 6 x^3 \ln (x) \over
12 (x-1)^5} . 
\end{eqnarray}

\subsection{${\cal L}_{\rm eff} (2l-2q)$ for $\Delta L_i = 1$}

In order to study the $\mu^- +$ Ti $\rightarrow e^- +$ Ti, and  
$\tau \rightarrow \mu ({\rm or}~ e) $ + (neutral meson),  
we need the effective Lagrangian for $l_i + q \rightarrow l_j + q$ where $q$
denotes a specific quark flavor. From Feynman diagrams analogous to Figs.~1 
and 2, we obtain 
\begin{equation}
{\cal L}_{\rm penguin}^{2l-2q}  = - 
{ 2 \alpha^2 \over m_{\tilde{l}}^2}~\left( \delta_{ji}^l \right)_{LL} 
P_1 (x) \overline{l_j} \gamma^{\mu} P_L l_i ~\sum_{q=u,d,s} e_q \overline{q} 
\gamma_{\mu} q + (L\rightarrow R), 
\end{equation}
\begin{equation}
{\cal L}^{2l-2q}_{\rm box} = \frac{-4  \alpha^2 }{m_{\tilde{l}}^2}
\left[ \left( \delta_{ji}^l \right)_{LL} \left(2 B_1(x) + B_2(x)
\right)~\overline{l_j}\gamma^{\mu}P_L l_i ~\sum_{q=u,d,s} e_q^2 
\overline{q} \gamma_{\mu} (P_L - P_R ) q - (L\leftrightarrow R) \right]
\end{equation}
where we assume $m_{\tilde{l}}=m_{\tilde{q}}$ for simplicity \footnote{  
In general, the function $P_1, B_1$ and $B_2$ should be generalized as 
functions of two variables, $x \equiv m_{\tilde{\gamma}}^2 / m_{\tilde{l}}^2$
and $y \equiv m_{\tilde{q}}^2 / m_{\tilde{l}}^2$ because of the difference 
between the slepton and squark masses.}.   
Again the functions $P_i$'s and $B_i$'s are originated from the penguin and
the box diagrams, respectively.  Note that ${\cal L}_{\rm penguin}^{2l-2q}$
and ${\cal L}^{2l-2q}_{\rm box}$ can be obtained from  Eqs.~(1)--(3) by 
replacing $e_l^4 \rightarrow e_l^3 e_q$ and $e_l^4 \rightarrow e_l^2 e_q^2$, 
respectively. The penguin contribution ${\cal L}_{\rm penguin}^{2l-2q}$ 
contains the vector quark current, and thus can contribute to $\mu^- +$ Ti 
$\rightarrow e^- +$ Ti, and $\tau \rightarrow l_{i\neq 3} + V (\equiv \rho^0, 
\phi)$. On the other hand, the box contribution ${\cal L}_{\rm 
box}^{2l-2q}$ depends only on the axial vector quark current so that it 
cannot contribute to the aformentoned processes, but it is relevant  to the 
process $\tau \rightarrow l_{i\neq 3} + P (\equiv \pi^0, \eta)$. One also 
has to include the operator $O_7$ describing $l_i \rightarrow l_j + \gamma$ 
to the above effective Hamiltonian when calculating physical amplitude for 
$2l-2q$ processes.  

\subsection{${\cal L}_{\rm eff} $ for $\Delta L_i = 2$}

In this subsection, we derive the effective Lagrangian for $\Delta L_i = 2$
for completeness. This Lagrangian is relevant to the muonium $\rightarrow$ 
antimuonium conversion, although the resulting effect turns out to be 
too small. The relevant Feynman diagrams are shown in Fig.~3. 
The results are 
(we fix $i=1, j=2$ in this subsection)
\begin{equation}
{\cal L}_{\rm eff}^{\Delta L_i = 2} = \sum_{i=1}^5 C_i^{\Delta L_i = 2} Q_i,
\end{equation}
where the basis operators in the effective theory are defined as

\begin{eqnarray}
Q_1 & = & \overline{e} \gamma_{\alpha} P_L \mu ~\overline{e} \gamma^{\alpha}
P_L \mu,
\nonumber   \\
Q_2  & = & \overline{e} \gamma_{\alpha} P_R \mu ~\overline{e} \gamma^{\alpha}
P_R \mu,
\nonumber   \\
Q_3 & = &  \overline{e} P_R \mu ~\overline{e} P_R \mu,
\nonumber   \\
Q_4 & = &  \overline{e} P_L \mu ~\overline{e} P_L \mu,
\nonumber   \\
Q_5 & = & \overline{e} P_L \mu ~\overline{e} P_R \mu.
\end{eqnarray}

\noindent
By matching the full theory amplitude with the effective amplitude, one can
obtain the Wilson coefficients as follow :

\begin{eqnarray}
C_1^{\Delta L_i = 2} 
& = & {\alpha^2 \over m_{\tilde{l}}^2}~\left( \delta_{12}^{l} 
\right)_{LL}^2 ~\left\{ {1\over 2} \tilde{f}_{6} (x) + x f_{6} (x) \right\},  
\nonumber    \\
C_2^{\Delta L_i = 2} & = & C_1 ~( {\rm with } ~L\leftrightarrow R),
\nonumber    \\
C_3^{\Delta L_i = 2}
 & = & - {\alpha^2 \over m_{\tilde{l}}^2}~ \left( \delta_{12}^l 
\right)_{LR}^2 ~2 x f_6 (x),
\nonumber    \\
C_4^{\Delta L_i = 2} & = & C_3 ~({\rm with} ~L \leftrightarrow R),
\nonumber    \\
C_5^{ \Delta L_i = 2} 
& = & { \alpha^2 \over m_{\tilde{l}}^2}~\left\{ \left( \delta_{12}^l 
\right)_{LL} \left( \delta_{12}^l \right)_{RR} ~\left[ 2 \tilde{f}_6 (x) + 
4 x f_6 (x) \right] - \left( \delta_{12}^l \right)_{LR} \left( \delta_{12}^l
\right)_{RL} ~4 \tilde{f}_6 (x) \right\}.
\end{eqnarray} 

\noindent
Here, the functions $\tilde{f}_6 (x)$ and $f_6 (x)$ are defined as
\begin{eqnarray}
\tilde{f}_6 (x) & = & {6 x (1 + 3x ) \ln x - x^3 - 9 x^2 + 9 x + 1 
\over 3 (x-1)^5},
\nonumber   \\
f_6 (x) & = & {6 (1 + 3 x) \ln x + x^3 - 9x^2 - 9x + 17 \over 6 (x-1)^5}.
\end{eqnarray} 
This completes the derivation of effective Lagrangians for $\Delta L_i = 1$ 
and 2 that will be used in the following sections.  
Also, for the purpose of $Z \rightarrow l_i \overline{l}_{j\neq i}$, we 
present the amplitude for this decay. In this case, we need the full 
amplitude as given in Sec.~III E. 

\section{Analytic expressions and numerical analyses \\ 
for various LFV processes}

\subsection{$l_i \rightarrow l_j + \gamma$}

The amplitude for $l_i \rightarrow l_j + \gamma^*$  can be written as 
\begin{equation}
{\cal M} (l_i^- \rightarrow l_j^- +\gamma^*) =  e \epsilon^{\alpha *}
\overline{u}_j (p-q) \left[ q^2 \gamma_{\alpha} \left( A_1^L P_L + A_1^R 
P_R \right) + i m_i \sigma_{\alpha\beta} q^{\beta} \left( A_2^L P_L + 
A_2^R P_R \right) \right] u_i (p)
\end{equation}
with
\begin{eqnarray}
A_1^L & = & - {\alpha \over 2 \pi m_{\tilde{l}}^2}~\left( \delta_{ji}^l 
\right)_{LL} P_1 (x), 
\nonumber   \\
A_1^R & = & A_1^L ({\rm with}~L\leftrightarrow R),
\nonumber   \\
A_2^R & = & -{C_7 \over 4 \pi^2} = - {\alpha \over 2 \pi m_{\tilde{l}}^2}~
\left[ M_3 (x) \left( \delta_{ji}^l \right)_{LL} 
+ {m_{\tilde{\gamma}} \over m_{l_i}} M_1 (x) \left( \delta_{ji}^l 
\right)_{LR} \right],
\nonumber \\ 
A_2^L & = &  A_2^R ({\rm with}~L\leftrightarrow R).
\end{eqnarray}
The decay rate for $l_i \rightarrow l_j + \gamma$ is 
\begin{equation}
\Gamma ( l_i \rightarrow l_j + \gamma ) = {\alpha \over 4}~m_i^5~
\left( | A_2^L |^2 + | A_2^ R |^2 \right). 
\end{equation}
Note that only the transition magnetic form factors $A_2^{L,R}$ contribute
to the on--shell photon emission.  The off--shell photon contribution 
($A_{1}^{L,R}$ form factors) is relevant to the $\mu \rightarrow 3 e$ and 
$\mu^- +$ Ti $\rightarrow e^- +$ Ti.
Normalizing it to the decay rate for $l_i \rightarrow l_j \nu_i 
\overline{\nu_j}$, one gets 
\begin{eqnarray}
B(  l_i \rightarrow l_j + \gamma ) & = & {\alpha^3 \over G_F^2}{12 \pi \over
m_{\tilde{l}}^4}~\left\{ \left| M_3 (x) \left( \delta_{ji}^l \right)_{LL} 
+ {m_{\tilde{\gamma}} \over m_{l_i}} M_1 (x) \left( \delta_{ji}^l \right)_{LR}
\right|^2 + (L \leftrightarrow R) \right\}
\nonumber    \\
& \times & B( l_i \rightarrow l_j \nu_i \overline{\nu_j} ).
\end{eqnarray}
One can derive the limits on $\delta_{ij}^l$'s from the experimental upper 
bounds listed in Table I, 
assuming there is no fortuitous cancellations among various terms, 
as in Ref.~\cite{masiero}, see Table II. 
Without such assumption, one 
would get a band in the $(  (\delta_{ij}^l)_{LL}, (\delta_{ij}^l)_{LR} )$ 
plane (see the solid lines in Fig.~4).

\subsection{$\mu^- + {\rm Ti} \rightarrow e^- + {\rm Ti}$}

The muon conversion to an electron on the titanium target is one of the
most sensitive probes of LFV that may arise from physics beyond the SM.  
In our case, the transition amplitude for this process can be expressed as
\begin{eqnarray}
{\cal M} (\mu^- + {\rm Ti} \rightarrow e^- + {\rm Ti}) 
& = & - {e^2 \over q^2}~\overline{e} \left[ q^2 \gamma^{\alpha} 
\left( A_1^L P_L + A_1^R P_R \right)  \right. \\
& + & \left. m_{\mu} i \sigma^{\alpha \beta} q_{\beta} \left( A_2^L P_L 
+ A_2^R P_R \right) \right] \mu \times \sum_{q=u,d} e_q \overline{q} 
\gamma_{\alpha} q,   \nonumber  
\end{eqnarray}
where $A_{1,2}^{L,R}$'s are defined in Eq.~(12). Note that there is no box 
contribution to this process, since only the quark vector current is 
important for the cohenrent conversion on the Ti nucleus. Also as alluded 
before, there is no $Z-$penguin contribution to this process to the order 
we are working.

The transition rate is given by 
\begin{equation}
\Gamma (\mu^- + {\rm Ti} \rightarrow e^- + {\rm Ti}) = 4 \alpha^5 
{Z_{\rm eff}^4 \over Z}~Z^2~| F ( q^2 \simeq - m_{\mu}^2 ) |^2 m_{\mu}^5 
\left[ |  A_1^L - A_2^R  |^2 + | A_1^R - A_2^L |^2 \right],
\end{equation}

\noindent For the Titanium target, $Z = 22, A = 48, N = 26, Z_{\rm eff} = 
17.6$ and $| F ( q^2 \simeq -m_{\mu}^2 ) | \simeq 0.54.$   
The experimental limit on the transition rate is given by
\begin{equation}
\Gamma ( \mu^- + {\rm Ti} \rightarrow e^- + {\rm Ti} ) < 
6.1 \times 10^{-13} ~\Gamma ( \mu ~{\rm capture~ in~ Ti} ),
\end{equation}
where the muon capture rate in Ti is $\Gamma (\mu ~{\rm capture ~in~ Ti} ) =
(2.590 \pm 0.012 ) \times 10^{6}/$  sec. Thus one gets the following upper 
bound : 
\begin{equation}
 |  A_1^L - A_2^R  |^2 + | A_1^R - A_2^L |^2 < \left[ 4.0 \times 10^{-11} 
{\rm GeV}^{-2} \right]^2.
\end{equation}
Thus, we get 
\begin{equation}
{1 \over m_{\tilde{l}}^2}~
\left| \left( \delta_{12}^l \right)_{LL} \left( - P_1 (x) + M_3 (x) \right) 
+ {m_{\tilde{\gamma}} \over m_{\mu}} \left( \delta_{12}^l \right)_{LR} 
M_1 (x) \right| < 3.4 \times 10^{-8}~{\rm GeV}^{-2} ,
\end{equation}
and similarly for the $(L\leftrightarrow R)$ case. This is another strong 
constraint that is independent of that from $\mu \rightarrow e \gamma$.  

As we noted in the previous subsection, without assuming that there is no
fortuitous cancellations among various terms, one would get a band in the
in the $((\delta_{ij}^l)_{LL}, (\delta_{ij}^l)_{LR} )$. One can not constrain
$(\delta_{ij}^l)_{LL}$ and $(\delta_{ij}^l)_{LR}$ independently of each other
without any assumptions. 
But, combing two different experiments, as one see in Fig.~4, 
we obtain two different bands from $\mu\rightarrow e\gamma$ and
$\mu^- + {\rm Ti} \rightarrow e^- + {\rm Ti}$. Only the shaded region
are allowed. 

\subsection{$l_i \rightarrow 3 l_j$ and $l_i \rightarrow l_{j \neq k} 
l_k \bar{l}_k$}

The amplitude for $l_i^- \rightarrow l_j^- l_j^+ l_j^- (\equiv 3 l_j)$ 
can be calculated from the effective Lagrangians, Eqs.~(1)--(3). It can be
written as the sum of the electromagnetic penguin and the box contributions :
\begin{eqnarray}
{\cal M}_{\rm penguin} & =  & \overline{u_j} (p_1) \left[ q^2 \gamma^{\alpha} 
\left( A_1^L P_L + A_1^R P_R \right) + m_{l_j} i \sigma^{\alpha\beta} 
q_{\beta} \left( A_2^L P_L + A_2^R P_R \right) \right] u_i (p)
\nonumber   \\
& \times & {e^2 \over q^2}~\overline{u_j} (p_2) \gamma_{\alpha} v_j (p_3) 
- ( p_1 \leftrightarrow p_2 ) 
\\
{\cal M}_{\rm box} & = & B_1^L e^2 \overline{u_j} (p_1) \gamma^{\alpha} 
P_L u_i (p)~\overline{u_j} (p_2) \gamma_{\alpha}P_L v_j (p_3)
\nonumber   \\
& + & B_1^R e^2 \overline{u_j} (p_1) \gamma^{\alpha} 
P_R u_i (p)~\overline{u_j} (p_2) \gamma_{\alpha} P_R v_j (p_3)
\nonumber \\
& + &  B_2^L e^2 \left[ \overline{u_j} (p_1) \gamma^{\alpha} 
P_L u_i (p)~\overline{u_j} (p_2)  \gamma_{\alpha}P_R v_j (p_3) - (p_1 \leftrightarrow p_2 ) 
\right]
\nonumber   \\
& + & B_2^R e^2 \left[ \overline{u_j} (p_1) \gamma^{\alpha} 
P_R u_i (p)~\overline{u_j} (p_2) \gamma_{\alpha} P_L v_j (p_3)
 - (p_1 \leftrightarrow p_2 ) 
\right].
\end{eqnarray}
Here the box form factors $B$'s are given by 
\begin{eqnarray}
B_1^L & = & - {2 \alpha \over \pi m_{\tilde{l}}^2} \left( \delta_{ji}^l 
\right)_{LL} ~( 2 B_1 (x)  + B_2 (x) )
\nonumber   \\
B_2^L & = & - {1\over 2}~B_1^L,
\nonumber   \\
B_1^R & = & B_1^L ({\rm with}~L\leftrightarrow R),
\nonumber   \\
B_2^R & = & B_2^L ({\rm with}~L\leftrightarrow R).
\end{eqnarray}

The decay rate for $l_i^- \rightarrow l_j^- l_j^- l_j^+$ is 
\begin{eqnarray}
\Gamma (l_i^- \rightarrow l_j^- l_j^- l_j^+) & = & {\alpha^2 \over 32 \pi}~
m_{l_i}^5~\left[ | A_1^L|^2 + | A_1^R |^2 - 2 ( A_1^L A_2^{R*} + A_2^L
A_1^{R*} + h.c.) \right.    
\nonumber   \\ 
& & + ( | A_2^L |^2 + | A_2^R |^2 ) \left( {16 \over 3} \ln {m_{l_i} \over
2 m_{l_j}} - {14 \over 9} \right)
\nonumber  \\
& & + {1\over 6} ( | B_1^L |^2 +  | B_1^R |^2 ) + 
{1 \over 3} ( | B_2^L |^2 + | B_2^R |^2 ) 
\nonumber \\
& & + {1 \over 3}~( A_1^L B_1^{L*} + A_1^L B_2^{L*} + 
 A_1^R B_1^{R*} + A_1^R B_2^{R*} + h.c. )
\nonumber  \\
& &  \left. - {2 \over 3}~( A_2^R B_1^{L*} + A_2^R B_2^{L*} + 
 A_2^L B_1^{R*} + A_2^L B_2^{R*} + h.c. ) 
\right].   
\end{eqnarray}

In case $j \neq k$, one has to remove the term with $p_1 \leftrightarrow 
p_2$ from the above amplitude and divide the $B_1^L$ and $B_1^R$ terms by 
a factor of 2.  Then,  the decay rate becomes
\begin{eqnarray}
\Gamma (l_i^- \rightarrow l_j^- l_k^- l_k^+) & = & {\alpha^2 \over 48 \pi}~
m_{l_i}^5~\left[ | A_1^L|^2 + | A_1^R |^2 - 2 ( A_1^L A_2^{R*} + A_2^L
A_1^{R*} + h.c.) \right.    
\nonumber   \\ 
& & + ( | A_2^L |^2 + | A_2^R |^2 ) \left( 8 \ln {m_{l_i} \over
m_{l_k}} - 12 \right)
\nonumber  \\
& & + {1\over 8} ( | B_1^L |^2 +  | B_1^R |^2 ) + 
{1 \over 2} ( | B_2^L |^2 + | B_2^R |^2 ) 
\nonumber \\
& & + {1 \over 4}~( A_1^L B_1^{L*} + A_1^R B_1^{R*} + h.c. )
\nonumber  \\
& & + {1 \over 2}~( A_1^L B_2^{L*} + A_1^R B_2^{R*} + h.c. )
\nonumber  \\
& & - {1 \over 2}~( A_2^R B_1^{L*} + A_2^L B_1^{R*} + h.c. ) 
\nonumber \\
& &  \left. -~( A_2^R B_2^{L*} + A_2^L B_2^{R*} + h.c. ) 
\right].   
\end{eqnarray}

We calculate the branching ratios for $\mu\rightarrow e\gamma$ and 
$\mu\rightarrow 3e$ in the allowed region shown in Fig.~4 for $x=0.3$
$0.9$, and $3.0$ assuming $\delta$'s are real. 
For $x=0.3$, the decay rate for $\mu\rightarrow 3e$ is 
dominated by the term which is propotional to 
$( | A_2^L |^2 + | A_2^R |^2 )$, namely 
$\mu\rightarrow e\gamma \rightarrow 3 e$. So, there is a strong correlation 
between the decay rates for $\mu\rightarrow e\gamma$ and
$\mu\rightarrow 3e$, see Eq.~(13) and Fig.~5~(a). The solid line in Fig.~5~(a) 
denotes this correlation. 
For larger $x$, this correlation becomes weaker and disappears.
>From Fig.~5, we observe that the branching ratio for
$\mu\rightarrow e\gamma \rightarrow 3 e$ is smaller than the present upper
bounds for $x<1$ and the region of high 
$B(\mu\rightarrow e\gamma)$ and low
$B(\mu\rightarrow e\gamma \rightarrow 3 e)$ is not allowed 
in the models under considerations.

Similar analyses could be done for $\tau\rightarrow 3e$
and $\tau\rightarrow 3\mu$. In the case of $\tau$ decay, there are no
independent experiments like as
$\mu^- + {\rm Ti} \rightarrow e^- + {\rm Ti}$ for 
$\mu\rightarrow e\gamma$ decay at present. So, one could not good predictions
for $\tau$ decays at present. (See the discussions below Eq.~(29).)

\subsection{$\tau \rightarrow l_{i=1,2} + $(neutral meson)}  

In this subsection, we consider the LFV in tau decays into a lighter 
lepton ($e$ or $\mu$) plus a light meson 
such as $\pi^0, \eta$ and $\rho^0$.
Different decays depend on different form factors so that each 
decay mode deserve its own study.  Because of the limited numbers of tau 
leptons that have been accumulated upto now, the typical upper limits on the 
branching ratios of LFV tau decays are order of $\sim 10^{-6}$.  
The limits on LFV tau decays  may be improved in the 
future at Tau-Charm factories or B factories. Therefore, it is important to 
study every possible LFV tau decays in various LFV models beyond the SM. In 
this subsection, we consider tau decays into a lighter lepton ($e$ or $\mu$) 
plus one light meson such as $\pi^0, \eta, \rho^0$ or $\phi$.  
 
The amplitude for $\tau \rightarrow l_{1=1,2} +$ (neutral pseudoscalar meson 
$\equiv P$ such as $\pi^0, \eta$, etc.) can be derived from the effective 
lagrangian (6) induced by the box diagrams :
\begin{eqnarray}
{\cal M} ( \tau (k,s) \rightarrow l_i (k^{'},s^{'}) + P (p) ) & = & 
 { 4 \alpha^2 \over m_{\tilde{l}}^2}~\left( \delta_{i3}^l 
\right)_{LL}~\left( 2 B_1 (x) + B_2 (x) \right)~
\nonumber \\
& \times & \overline{l}_i \gamma^{\alpha} P_L \tau ~ \sum_{q} e_q^2~
\langle P(p) | \overline{q} \gamma_{\alpha} \gamma_5 q | 0 \rangle + 
(L\rightarrow R).    
\end{eqnarray}
Using the PCAC relations, and assuming that $| \eta \rangle = | ( u \bar{u} + 
d \bar{d} - 2 s \bar{s} ) / \sqrt{6} \rangle$, one gets 
\begin{equation}
\langle P (p) | \sum_q e_q^2 \overline{q} \gamma_{\alpha} \gamma_5 q | 0 
\rangle = i {1 \over 3}~C_P~f_{\pi} p_{\alpha},
\end{equation}
with $C_{\pi^0} = 1, C_{\eta} = 1/\sqrt{3}$ and $f_{\pi} = 93$ MeV.
Then the decay rate for this decay is given by
\begin{equation}
\Gamma ( \tau \rightarrow l_i + P ) 
= {m_{\tau}^3 f_{\pi}^2 \over 18 \pi}~{\alpha^4 \over m_{\tilde{l}}^4}~
\left( 2 B_1 + B_2 \right)^2 C_P^2~
\left\{ \left| ~ \left( \delta_{i3}^l \right)_{LL} \right|^2 + \left|~ 
\left( \delta_{i3}^l \right)_{RR} \right|^2 \right\}.
\end{equation}
The decay rate depends on $(\delta_{i3})^l_{LL}$ and $(\delta_{i3})^l_{RR}$, 
it does not depend on $\delta_{LR,RL}$.

The amplitudes for $\tau \rightarrow l_{i=1,2} + $ (neutral vector meson 
$\equiv V$ such as $\rho^0, \phi$) can be calculated using the 
effective lagrangian (1), (5) induced by the penguin diagrams : 
\begin{equation}
{\cal M} ( \tau (k,s) \rightarrow l_i (k^{'},s^{'}) + V (p,\epsilon^{*}) )
 \equiv  { 2 \alpha^2 \over m_{\tilde{l}}^2 }~C_V f_V m_V~ 
\left[
~A_{L}^{\tau} \overline{l_i} \gamma^{\alpha} P_L \tau 
+ ( L \leftrightarrow R ) ~\right] \epsilon_{\alpha}^*,
\end{equation}
where 
\begin{equation}
A_{L}^{\tau}  =  \left( \delta_{i3}^l \right)_{LL} \left\{- P_1 (x) 
+ { m_{\tau}^2 \over m_V^2 }~M_3 (x) \right\} 
+ {m_{\tau}^2 \over m_V^2} \times {m_{\tilde{\gamma}} \over m_{\tau}}
~\left( \delta_{i3}^l \right)_{LR} M_1 (x),  
\end{equation}
and similarly for $A_R^{\tau}$. This decay is a complete analogue of $\mu^-$ 
+ Ti $\rightarrow e^-$ + Ti at the parton level.  So we expect that we can
constrain $\delta_{LL}$ and $\delta_{LR}$ without any assumptions combining
$\tau\rightarrow l+\gamma$ and $\tau\rightarrow l+V$ in the future.
When writing the amplitude 
in the above form, we have used the definition of the vector meson decay 
constant $f_V$ :
\begin{equation}
\sum_q \langle V | e_q \overline{q} \gamma^{\alpha} q | 0 \rangle =
C_V f_V m_V \epsilon^{\alpha *},  
\end{equation}
with $C_{\rho^0} =1, C_{\phi} = -1/3$ and
$f_{\rho^0}=153~{\rm MeV},~f_{\phi}=237~{\rm MeV}$.
The decay rate for $\tau \rightarrow l_i + V$ is 
\begin{equation}
\Gamma ( \tau \rightarrow l_i + V ) = 
\frac{f_V^2(m_{\tau}^2-m_V^2)(m_{\tau}^4+m_{\tau}^2 m_V^2-2m_V^4)}
{8\pi m_{\tau}^3}
\frac{\alpha^4}{m_{\tilde{l}}^4}
C_V^2\left\{  \left| A_{L}^{\tau} \right|^2+
\left| A_{R}^{\tau} \right|^2 \right\}.
\end{equation}
We calculate the branching ratios for $\tau\rightarrow l+P$ 
and $\tau\rightarrow l+V$ from the
constraints shown in Table II. The results are shown in Table III and IV.
Most of these decays are expected to occur with small branching ratios
far below  the current or future experimental search limits.
As such, it establishes the necessary amount of tau leptons in order to 
probe the LFV from nondiagonal slepton mass matrix. 

\subsection{$Z^0 \rightarrow l_i  \overline{l}_{j\neq i}$} 

The photino-mediated LFV can generate the LFV decays of $Z$ bosons.
The amplitude for $Z \to l_i l_j$ decays are given by
\bea
{\cal M} &=& - \left(\frac{\alpha}{2\pi}\right)
 \left(\frac{8 m_{_Z}^2 G_{_F}}{\sqrt{2}}\right)^{\frac{1}{2}}
 \frac{(v_l+a_l)}{2}
\nonumber \\
&& \times \left[ (\bar{u}(p')P_{_L}v(p)) 
                 \left( (\delta^l_{ij})_{LL} X^{\mu ij}_{LL}(z)
                       -(\delta^l_{ij})_{RL} X^{\mu ij}_{RL}(z)
                 \right)     
          \right.
\nonumber \\
&&\left. ~~~~~~~~~~~~~+ (\bar{u}(p')P_{_R} \gamma^{\mu}v(p)) 
                 \left( (\delta^l_{ij})_{LL} Y^{ij}_{LL}(z)
                       +(\delta^l_{ij})_{LR} Y^{ij}_{RL}(z)
                 \right)     
\right] \cdot \epsilon_{\mu} (Z) 
\nonumber \\
&& + ((v_l + a_l) \leftrightarrow (v_l-a_l),
~~ P_{_L} \leftrightarrow P_{_R} ),
\eea
where $z = m^2_{\tilde{\gamma}}/ m^2_{\tilde{l}}$ and
\bea
X^{\mu ij}_{LL}(z) &=& \frac{m_l}{m_{\tilde{l}}}
\left( \frac{q^\mu}{m_{\tilde{l}}} F_5(z)
     - \frac{p^\mu}{m_{\tilde{l}}} F_6(z)
\right),
\nonumber \\
X^{\mu ij}_{RL}(z) &=& \frac{m_{\tilde{\gamma}}}{m_{\tilde{l}}}
\left( \frac{q^\mu}{m_{\tilde{l}}} G_1(z)
     + \frac{p^\mu}{m_{\tilde{l}}} G_2(z)
\right),
\nonumber \\
Y^{ij}_{LL}(z) &=& 
 F_1(z) + \frac{m_l^2}{m^2_{\tilde{l}}} F_2(z)
-F_3(z) - \frac{m_l^2}{m^2_{\tilde{l}}} F_4(z),
\nonumber \\
Y^{ij}_{RL}(z) &=& 
\frac{m_{\tilde{\gamma}} m_l}
     {m^2_{\tilde{l}}} G_4(z),
\nonumber \\
\eea
with 
\begin{equation}
v_l = -\frac{1}{2} + 2 \sin^2 \theta_{_W},~~~~
a_l = -\frac{1}{2},
\end{equation}
and $m_l$ is the mass of the heavier lepton in the final state.
The functions $F_i(z)$ and $G_i(z)$ are defined as follows:
\bea
\frac{1}{m^2_{\tilde{l}}} F_1(z) &=& 
\int^1_0 dx ~\int^{1-x}_0 dy~
\frac{1-x}
{x m^2_{\tilde{\gamma}} + (1-x) m^2_{\tilde{l}} +y(x+y-1) m^2_{_Z}},
\nonumber \\
\frac{1}{m^4_{\tilde{l}}} F_2(z) &=& 
\int^1_0 dx ~\int^{1-x}_0 dy~
\frac{x(1-x)(1-x-y)}
{(x m^2_{\tilde{\gamma}} + (1-x) m^2_{\tilde{l}} +y(x+y-1) m^2_{_Z})^2},
\nonumber \\
\frac{1}{m^2_{\tilde{l}}} F_3(z) &=& 
\int^1_0 dx 
\frac{x^2}
{(1-x) m^2_{\tilde{\gamma}} + x m^2_{\tilde{l}}},
\nonumber \\
\frac{1}{m^4_{\tilde{l}}} F_4(z) &=& 
\int^1_0 dx 
\frac{x^3(1-x)}
{((1-x) m^2_{\tilde{\gamma}} + x m^2_{\tilde{l}})^2},
\nonumber \\
\frac{1}{m^4_{\tilde{l}}} F_5(z) &=& 
\int^1_0 dx ~\int^{1-x}_0 dy~
\frac{(1-x)(1-x-y)(1-2y)}
{(x m^2_{\tilde{\gamma}} + (1-x) m^2_{\tilde{l}} +y(x+y-1) m^2_{_Z})^2},
\nonumber \\
\frac{1}{m^4_{\tilde{l}}} F_6(z) &=& 
\int^1_0 dx ~\int^{1-x}_0 dy~
\frac{2x(1-x)(1-x-y)}
{(x m^2_{\tilde{\gamma}} + (1-x) m^2_{\tilde{l}} +y(x+y-1) m^2_{_Z})^2},
\nonumber \\
\eea
and
\bea
\frac{1}{m^4_{\tilde{l}}} G_1(z) &=& 
\int^1_0 dx ~\int^{1-x}_0 dy~
\frac{(1-x)(1-2y)}
{(x m^2_{\tilde{\gamma}} + (1-x) m^2_{\tilde{l}} +y(x+y-1) m^2_{_Z})^2},
\nonumber \\
\frac{1}{m^4_{\tilde{l}}} G_2(z) &=& 
\int^1_0 dx ~\int^{1-x}_0 dy~
\frac{-2x (1-x)}
{(x m^2_{\tilde{\gamma}} + (1-x) m^2_{\tilde{l}} +y(x+y-1) m^2_{_Z})^2},
\nonumber \\
G_3(z) 
&=& \frac{-1+z-\log z}{(1-z)^2},
\nonumber \\
G_4(z) 
&=& \frac{1+4z-5z^2+4z\log z+2z^2 \log z}{2(1-z)^4} = M_1(z).
\nonumber \\
\eea

The branching ratio of $Z \to l_i l_{j \neq i}$ processes is given by
\bea
\mbox{Br}(Z \to l_i l_{j \neq i}) &=& \mbox{Br}(Z \to e^+ e^-)
             \frac{\Gamma(Z \to l_i l_j)}{ \Gamma(Z \to e^+ e^-)}
\nonumber \\
&=& \mbox{Br}(Z \to e^+ e^-) \left( \frac{\alpha}{2\pi} \right)^2
    \frac{1}{2(v_l^2+a_l^2)}
\nonumber \\
&&~ \times \left[ (v_l+a_l)^2 
           \left( 
      | (\delta^l_{ij})_{LL} |^2 (F_1(z)-F_3(z))^2
     +| (\delta^l_{ij})_{LR} |^2 \frac{ m^2_{\tilde{\gamma}} m^2_{_Z}}
                                  { 8 m^4_{\tilde{l}}} G_2^2(z)
           \right) \right.               
\nonumber \\
&&~~~~~~  \left. (v_l-a_l)^2 
           \left( 
      | (\delta^l_{ij})_{RR} |^2 (F_1(z)-F_3(z))^2
     +| (\delta^l_{ij})_{RL} |^2 \frac{ m^2_{\tilde{\gamma}} m^2_{_Z}}
                                  { 8 m^4_{\tilde{l}}} G_2^2(z)
           \right) \right] ,
\eea
where $B( Z^0 \rightarrow e^+ e^- ) = 3.366 \%$ and
\bea
\Gamma(Z \to e^+ e^-) = \frac{G_{_F} m_{_Z}^3}{12 \pi \sqrt{2}}
                        \cdot 2 (v_l^2+a_l^2)~.
\nonumber
\eea
The upper limits on the LFV $Z$ decays 
are \cite{klaus}
\begin{eqnarray}
  B_{\rm exp} (Z \rightarrow e \mu ) < 2.5 \times 10^{-6}, &&  
\nonumber  \\
B_{\rm exp} (Z \rightarrow e \tau ) < 7.3 \times 10^{-6}, && 
\nonumber   \\
B_{\rm exp} (Z \rightarrow \mu \tau ) < 1.0 \times 10^{-5}, && 
\end{eqnarray}
The associated constraints on $\delta^l$'s 
are so loose that they are useless. 
Using the constraints obtained in the previous subsections, 
we find that the upper limits on the branching ratios for the LFV $Z$ 
decays are less than $10^{-7(8)}$ for $Z\rightarrow \mu (e)+\tau$, 
and $10^{-10}$ 
for $Z\rightarrow e\mu$, 
which are far below the present experimental results. 
Any observations of LFV $Z$ decays with $B > 10^{-7}$ will indicate that the 
source of LFV should be different from the nondiagonal slepton mass matrix
elements.

\subsection{Muonium $\rightarrow$ antimuonium conversion}

Now let us consider the muonium $\rightarrow$ antimuonium conversion.
The current experimental upper limit on the transition probability in the 
external magnetic field $B_{\rm ext} = 0.1$ T is 
\cite{mac}
\begin{equation}
P_{\rm exp} ( M \rightarrow \overline{M} ) < 8.2 \times 10^{-11} 
~~~(90 \% {\rm C.L.}).
\end{equation}
When this process is described the following effective Lagrangian,
\begin{equation}
{\cal L}_{\rm eff}^{M\to \bar{M}}  = {G_{--} \over \sqrt{2}}~\left( 
\overline{e} \mu \right)_{V-A} \left( \overline{e} \mu \right)_{V-A}
+ {G_{+-} \over \sqrt{2}}~\left( \overline{e} \mu \right)_{V+A} \left( 
\overline{e} \mu \right)_{V-A},
\end{equation}
it is known that the effective couplings $G_{\mp -}$ are constrained as
\begin{equation}
G_{--} < 3.0 \times 10^{-3}~G_F,~~~~{\rm or} ~~~~ 
G_{+-} < 2.1 \times 10^{-3}~G_F,
\end{equation}
assuming only one of them is nonvanishing. 

Now we can derive the muonium ($M \equiv \mu^+ e^-$)$\rightarrow$ antimuonium 
($\overline{M} \equiv \mu^- e^+$) conversion induced by the $\Delta L_i = 2$ 
effective Lagrangian, (7).  Note that our model predicts that 
\begin{eqnarray}
G_{\mp -} & \sim & 
{\alpha^2 \over m_{\tilde{l}}^2}~\left( \delta_{12}^l \right)^2
\times \tilde{f_6} (x) ~~({\rm or}~x f_6 (x))   
\nonumber  \\
& \lesssim & 4.6 \times 10^{-4} ( \delta_{12}^l )^2 G_F,
\end{eqnarray}
because $x f_6$ and $ \tilde{f}_6(x) $ are far less than one. 
So the effect of the nondiagonal
slepton mass matrix on the muonium conversion 
is totally negligible. In other words, if one observes
the muonium $\rightarrow$ antimuonium conversion, it would imply that the
origin of the associated lepton flavor violation should be something different
from what we consider in this work, for example $R-$parity violation 
\cite{r-parity} or dilepton gauge boson, etc. \cite{hou}.
 
\section{Constraints on flavor conserving mass insertion from the lepton
anomalous magnetic moments}

In this section, we consider the limits on the flavor conserving mass 
insertion $\left( \delta_{ii}^l \right)_{LR}$ that are derivable from
the anomalous magnetic moments of leptons.  
In Ref.~\cite{masiero}, this quantity was constrained from the condition
$\Delta m_{l}^{\rm SUSY}  < m_{\rm phys}$, where the $\Delta m_{l}^{\rm 
SUSY}$ is calculated by  a flavor conserving mass insertion and is finite :
\begin{equation}
\Delta m_{l}^{\rm SUSY} = - {2 \alpha \over 4 \pi}~m_{\tilde{\gamma}}~
{\rm Re}~\left( \delta_{ii}^l \right)_{LR}~I(x),
\end{equation}   
where $I(x) = (-1+x-x \ln x )/(1-x)^2$.  The physical mass $m_{\rm phys}$ 
will be given by $m_{\rm bare} + \delta m_{\rm c.t.} + 
\Delta m_{l}^{\rm SUSY}$, where $\delta m_{\rm c.t.}$ is the mass 
renormalization counter term in the MSSM. However its finite part is 
arbitrary, and one has to assume that there is no large cancellation
between it and $\Delta m_{l}^{\rm SUSY} $ in order to make use of it.
In other words, renormalizable couplings cannot be calculated from the first 
principle without any ambiguity. Therefore, the condition that  
$\Delta m_{l}^{\rm SUSY}  < m_{\rm phys}$ may be a plausable assumption,
but it is by no means  on the firm ground like the constraints considered 
in the previous sections. On the other hand, the anomalous magnetic moment 
of a lepton is calculable in the SM and any other renormalizable field 
theories without any ambiguity. So it is  meaningful to require 
$a_{l}^{\rm SM} + a_{l}^{\rm SUSY} = a_{l}^{\rm exp}$, which we adopt 
in the following 
\footnote{ 
There can be also potentially important contributions from chargino-sneutrino 
loop (in addition to the heavier neutralinos-slepton loops), especially when 
charginos and sneutrinos are light \cite{carena}. }. 

The anomalous magnetic moment of a lepton $l$ ($\equiv a_l$) is defined as   
\begin{equation}
a_l \equiv \left( {g-2 \over 2} \right) = F_2 (0),
\end{equation}
where $F_2(q^2)$ is the magnetic form factor of a lepton :
\begin{equation}
{\cal M} (l_i (p,s) \rightarrow l_i (p^{'},s^{'}) + \gamma (q,\epsilon) ) 
= \bar{u (p^{'},s^{'})} \left[ F_1 (q^2) \gamma^{\mu} +  i F_2 (q^2) 
{\sigma^{\mu \nu} q_{\nu} \over 2 m_i} \right] u_i (p,s) \epsilon_{\mu}^{*} 
(q). 
\end{equation}

In order to derive the SUSY contribution to the anomalous magnetic moment, 
one cannot use the effective Lagrangian presented in Sec.~II A, since we 
assumed that the final lepton mass is negligible when we derived Eq.~(1).
One has to go back to the original expression for the $l_{i} \rightarrow 
l_{j} + \gamma$ with $m_i = m_j$. It is straightforward to show that the 
flavor-conserving mass insertion ($\delta_{ii}^l$) induces 
\begin{equation}
a_{l}^{\rm SUSY} = - {\alpha \over \pi}~{ m_i^2 \over m_{\tilde{l}}^2 }~
\left[ P_3 (x) + M_3 (x) \left( \delta_{ii}^l \right)_{LL} + 
{m_{\tilde{\gamma}} \over m_i}~\left( \delta_{ii}^l \right)_{LR} M_1 (x) 
\right],
\label{eq:al}
\end{equation}
where the function $P_3 (x)$ is given by 
\begin{equation}
P_3 (x) = {1 - 6 x + 3 x^2 + 2 x^3 - 6 x^2 \ln x \over 6 ( 1 - x)^4},
\end{equation}
and $x \equiv m_{\tilde{\gamma}}^2 / m_{\tilde{l}}^2$ as before, and 
$M_{1,3} (x)$'s are defined in Sec.~II. Here we have assumed that 
$( \delta_{ii}^l )_{LL} = ( \delta_{ii}^l )_{RR}$, and  
$( \delta_{ii}^l )_{LR} = ( \delta_{ii}^l )_{RL}$. 
The first term in Eq.~(\ref{eq:al}) arises from the slepton--photino loop 
without any mass insertion. There is also another term proportional to 
$m_i^2$ coming from one insertion of $\Delta_{ii}^l )_{LL}$. However, this 
is suppressed compared to the above by an additional factor of 
$m_i / m_{\tilde{l}}$, and thus can be safely neglected.  From this 
expression, one can get the constraint on $x$ and 
$( \delta_{ii}^l )_{LR}$ for a given value of $m_{\tilde{l}}$.    

The current status of the anomalous magnetic moments for $e$ and $\mu$ 
are as follow \cite{kinoshita}:
\begin{eqnarray}
\left.
\begin{array}{l}
a_e^{\rm SM} = 1 159 652 156.4 (1.2) \times 10^{-12} \\
a_e^{\rm exp} = 1 159 652 188.25 (4.24) \times 10^{-12}
\end{array}
\right\}
&&\rightarrow  a_e^{\rm new} = 31.85 (4.4) \times 10^{-12} 
\nonumber  \\
\left.
\begin{array}{l}
a_{\mu}^{SM} = 116 591 711 (94) \times 10^{-11} \\
a_{\mu}^{\rm exp} =  1 165 923 (8.5) \times 10^{-9} 
\end{array}
\right\}
&&\rightarrow a_{\mu}^{\rm new} = 5.9 (8.6) \times 10^{-9}
\end{eqnarray}
Here, the SM predictions for electron and muon anomalous magnetic moments
include the one loop electroweak corrections and the two loop leading log 
terms, as well as QED corrections including the hadronic vacuum polarization 
and the hadronic light-by-light scattering \cite{kinoshita}.
We ignored the tau anomalous magnetic moment here, since the experimental
value begins to probe the lowest order QED correction at the present.
The resulting constraints on $( \delta_{ii}^l )_{LR}$'s for $i=1,2$ are 
shown in Table~V. 
Comparing with the constraints obtained in Ref.~\cite{masiero},
our constraints are more reliable
and even stronger for the case of muon.

The imaginary part of the flavor conserving mass insertion is constrained by 
the electric dipole moment (EDM) of a lepton, as discussed in Ref.~
\cite{masiero}. The bound from electron EDM is very strong, 
\begin{equation}
\left|~ {\rm Im} \left( \delta_{11}^l \right)_{LR} \right| \sim
({\rm a ~few~}\times 10^{-7} ).
\end{equation}

\section{Conclusion}

In conclusion, we considered the LFV in general SUSY models, where the slepton
mass matrices are not diagonal in the basis where $l-\tilde{l}-\tilde{\gamma}$
is diagonal.  We worked in the mass insertion approximations in which 
$\left( \delta_{ij}^l \right)_{LR}$'s constitute the suitable parameters 
that characterize the strengths of LFV. There are strong constraints on some 
of these parameters from $l_i \rightarrow l_j + \gamma$ and 
$\mu^- + {\rm Ti} \rightarrow e^- + {\rm Ti}$. 
Using these constraints, we predict the upper limits on other LFV processes 
such as $l_i \rightarrow 3 l_j$, 
$\tau \rightarrow \mu ~({\rm or}~e) +$ (neutral meson), 
$Z \rightarrow l_i l_{j \neq i}$ and the muonium $\rightarrow$ 
antimuonium conversion.  

LFV processes considered in this work are sensitive probes of the slepton
mass matrices which are related with the SUSY breaking mechanism.  
Any positive LFV signal would herald the existence of some new physics beyond 
the SM, and if the predictions in this work are violated, then one has to 
think of another source of LFV other than that through the nondiagonal 
slepton mass matrices. 
In particular, if one imposes the constraints from $l_i \rightarrow l_j +
\gamma$ and $\mu^- + {\rm Ti} \rightarrow e^- + {\rm Ti}$, 
then the expected ranges for other 
LFV processes are well below the current limits and the level to be achieved
in the near future.  If some LFV processes are observed at rates higher than 
those predicted in this work, the source of LFV would not be likely to be
photino-mediated.  For example, presence of some $R-$parity violating 
couplings can lead to quantitatively different predictions 
\cite{r-parity} from those made in this work. 

The constraints on the flavor conserving mass insertions were derived from
the anomalous magnetic moments of leptons.  These bounds
are to be considered more sensible than those obtained from $\Delta m_{\rm 
SUSY} < m_{\rm exp}$, since the renormalizable couplings in the 
renormalizable field theories cannot be calculated from the given lagrangian.
Our constraints still imply that the diagonal slepton masses should be almost 
degenerate, especially  for the first two generations.  One has to speculate 
why this should be the case in general supersymmetric models.   

Finally, let us comment on our assumption that the LSP is a pure photino, 
and other neutralinos are heavy enough so that their effects might be 
ignored. In order to do more complete analyses for given neutralino spectra 
({\it i.e.} for given $M_1, M_2, \mu$ and $\tan\beta$), one can easily 
include the effects of all the neutralinos in principle, and do the 
similar analyses as presented in this work. This is possible, since the 
neutralino spectrum is independent of the slepton spectra. Our approach 
adopted in this work can be regarded as a first step to such complete 
analyses.  The qualitative features of our predictions would not change 
very much. In other words, our predictions are expected to be correct within 
an  order of magnitude.

\acknowledgements
The authors thank Dr. S.Y. Choi for illuminating discussions on the 
Feynman rules in the presence of Majorana neutrinos. 
This work is supported in part by KOSEF 
through Center for Theoretical Physics at Seoul National University, 
by KOSEF Contract No. 971-0201-002-2, 
by the Ministry of Education through the Basic Science Research Institute,
Contract No. BSRI-97-2418 (PK), and KAIST Center for Theoretical Physics and
Chemistry (KYL,PK).

%
%
\begin{figure}[ht]
\hspace*{-1.0 truein}
\psfig{figure=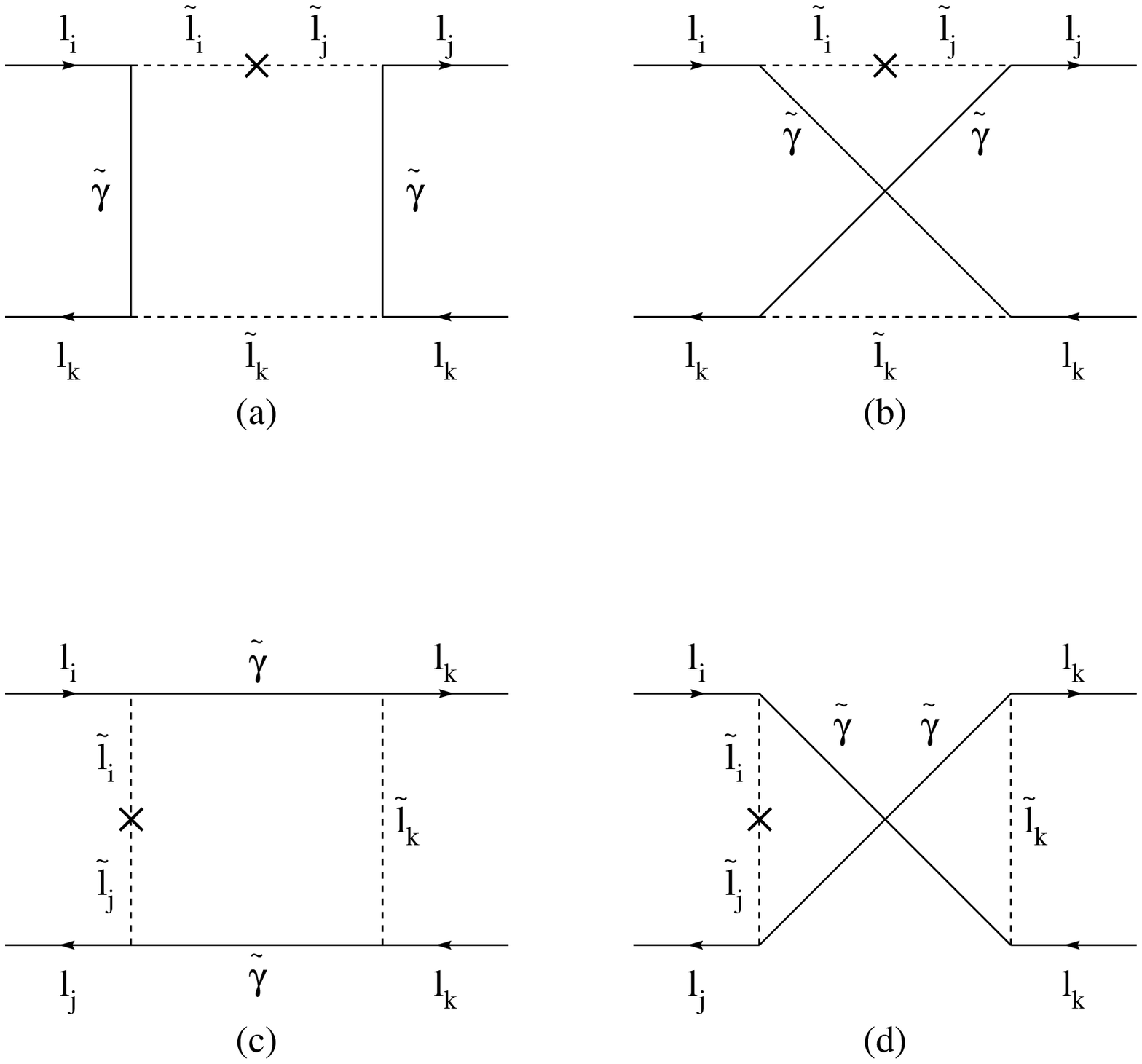}
\caption{Box diagrams for $\Delta L_i = 1$. 
Here $i,j,k,l$ are the generation indices. }
\label{fig1}
\end{figure}

\begin{figure}[ht]
\hspace*{-1.0 truein}
\psfig{figure=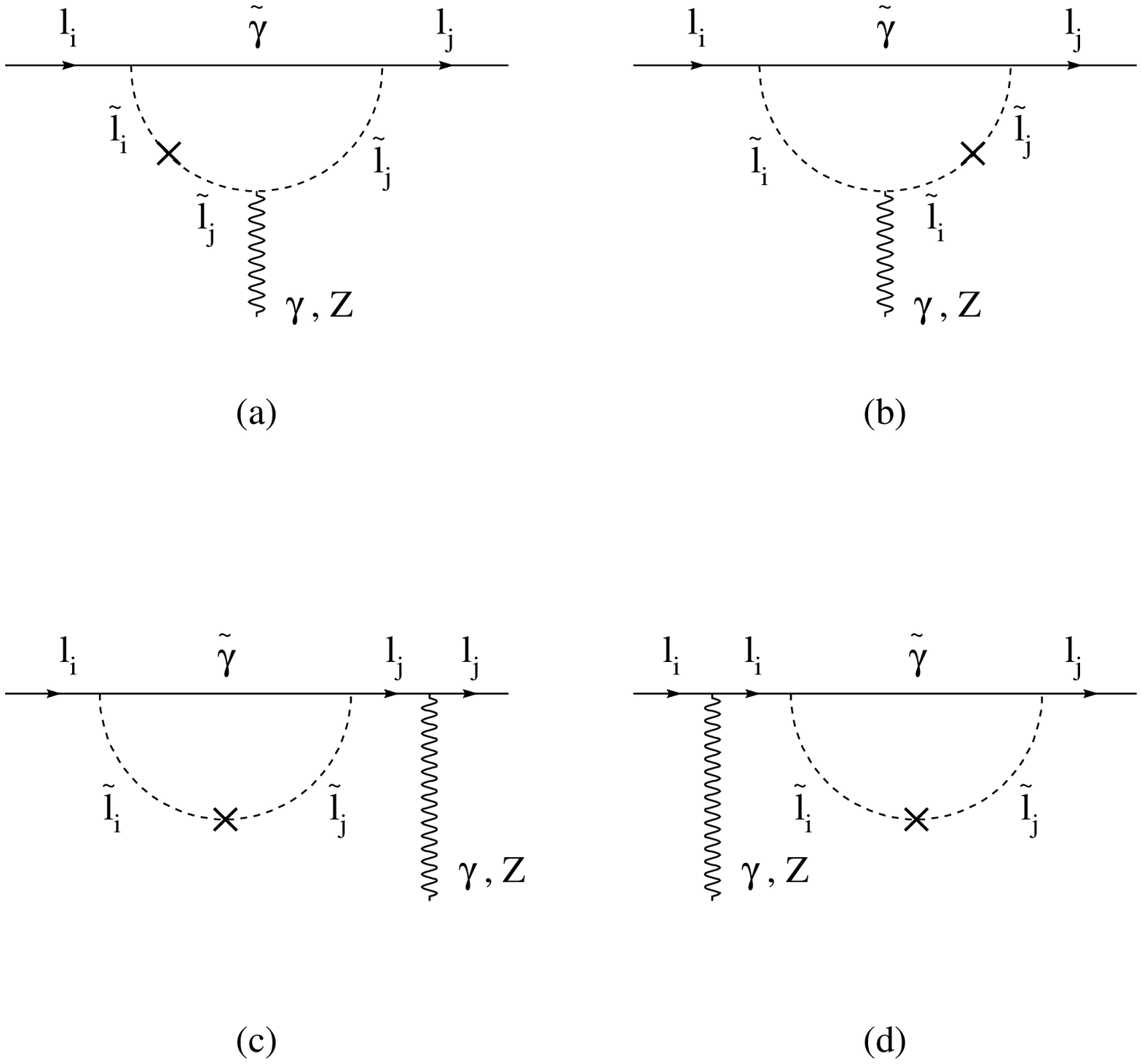}
\caption{Penguin diagrams for $\Delta L_i = 1$.
Here $i,j,k,l$ are the generation indices. }
\label{fig2}
\end{figure}

\begin{figure}[ht]
\hspace*{-1.0 truein}
\psfig{figure=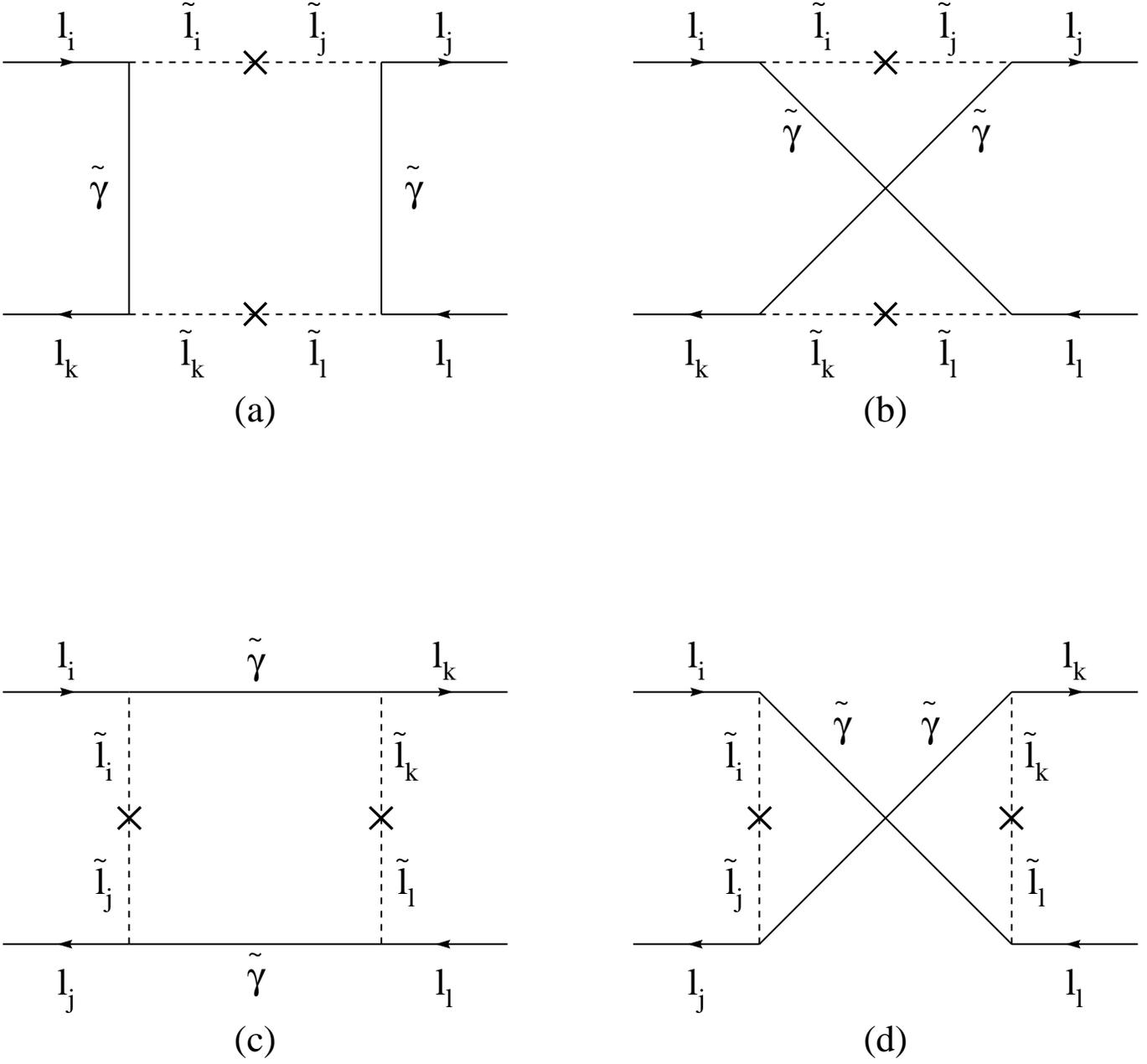}
\caption{Feynman diagrams for $\Delta L_i = 2$.
Here $i,j,k,l$ are the generation indices. }
\label{fig3}
\end{figure}

\begin{figure}[ht]
\hspace*{-1.0 truein}
\psfig{figure=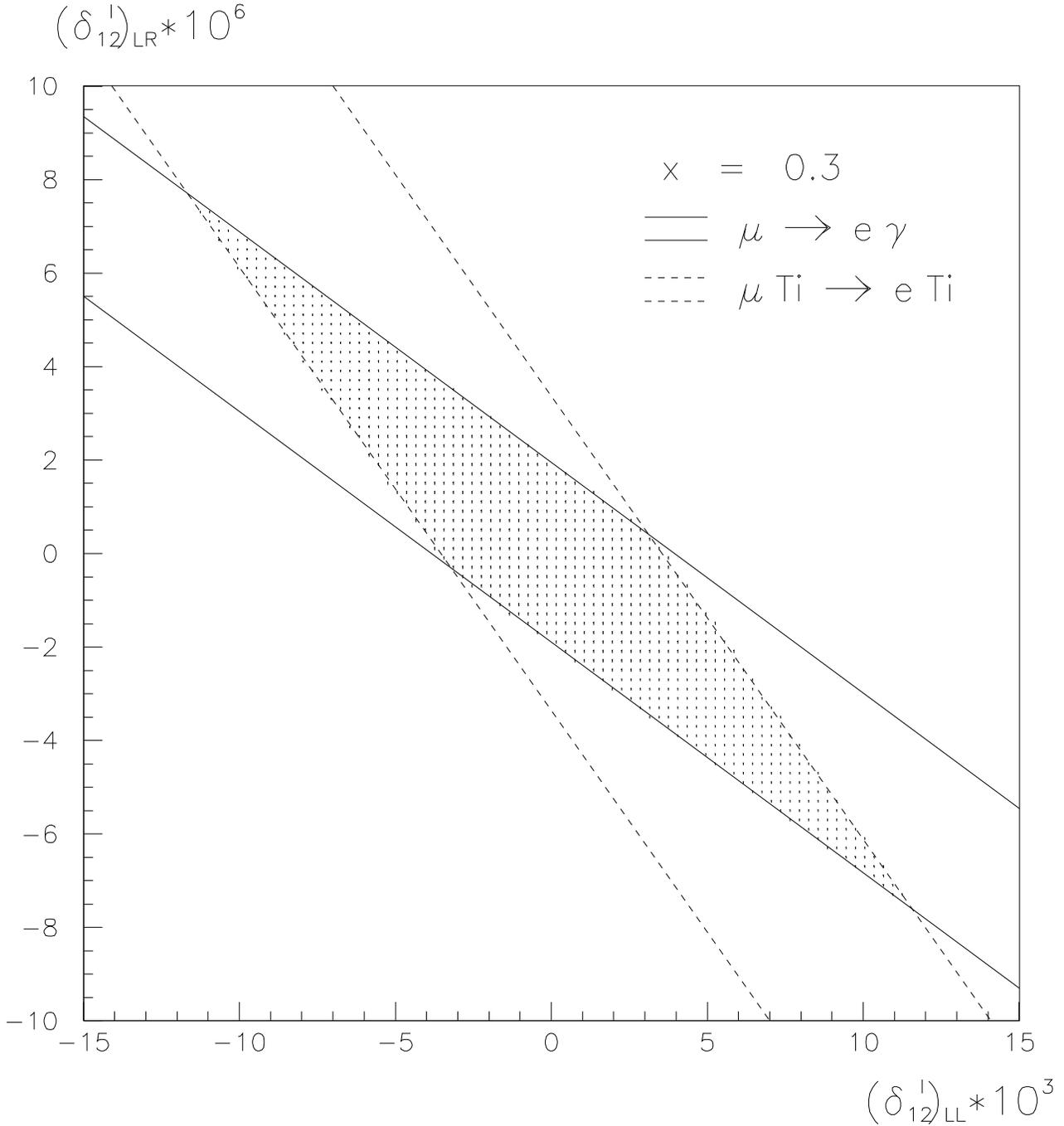}
\caption{Allowed regions in the $( ( \delta_{12}^l)_{LL},(\delta_{12}^l
)_{LR} )$ planes for $x=0.3$.  
The region between solid lines are allowed by
the present experiment for $\mu\rightarrow e\gamma$ and
the region between dotted lines are allowed by
the present experiment for $\mu^-~{\rm Ti} \rightarrow e^-~{\rm Ti}$.
Combining two experiments, only the shaded region is allowed.
}
\label{fig4}
\end{figure}

\begin{figure}[ht]
\hspace*{-1.0 truein}
\psfig{figure=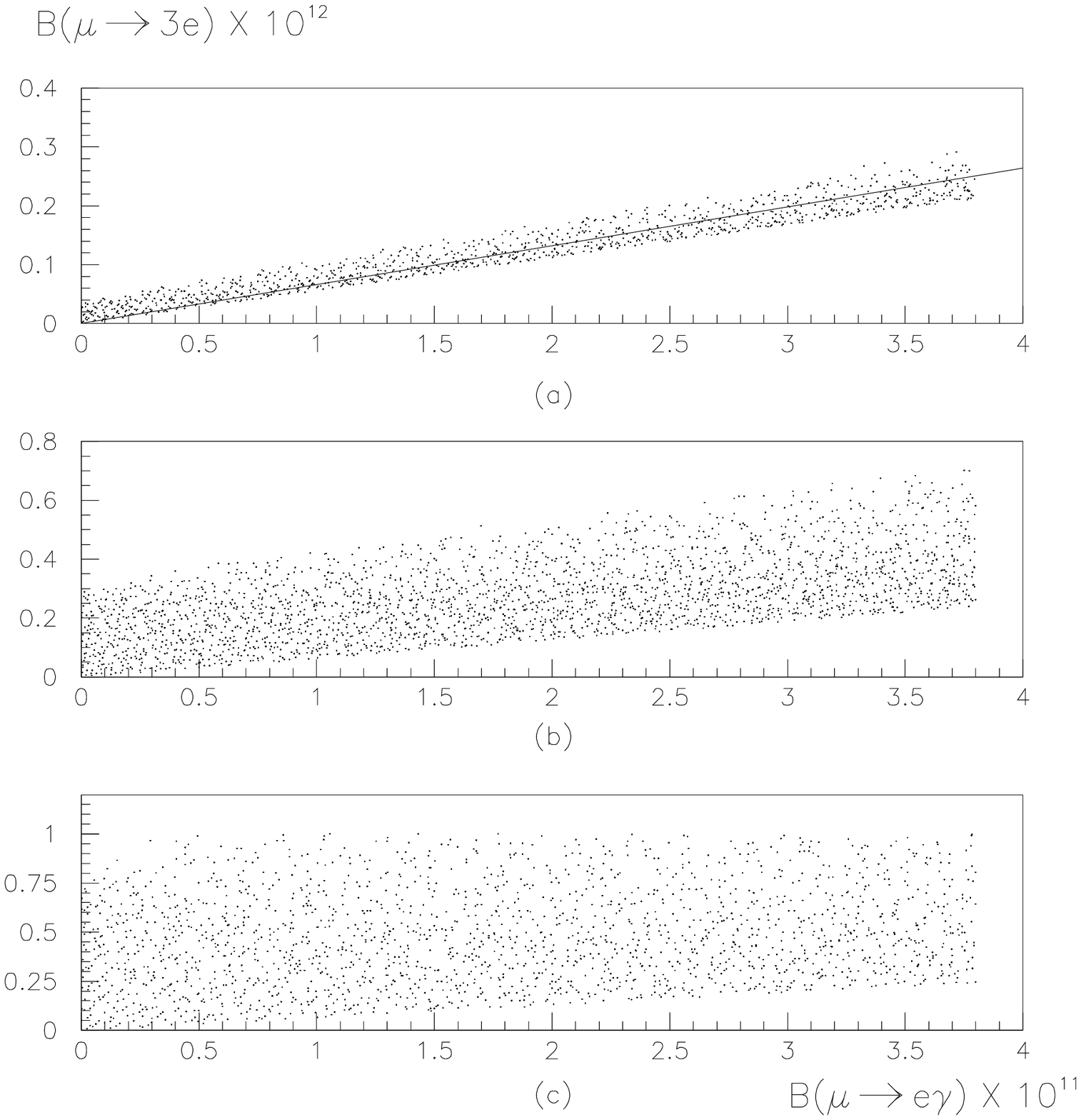}
\caption{
The branching ratios for $\mu\rightarrow e\gamma$ and
$\mu\rightarrow 3e$ in the allowd region 
by the present experiment for $\mu\rightarrow e\gamma$ and
$\mu^-~{\rm Ti} \rightarrow e^-~{\rm Ti}$
for $x=0.3$~(a), $x=0.9$~(b), and $x=3$~(c)
 assuming $\delta$'s are real.  
The solid line  in (a) denotes the correlation between
the decay rates for $\mu\rightarrow e\gamma$ and
$\mu\rightarrow 3e$. 
}
\label{fig5}
\end{figure}

\begin{table}
\caption{Upper limits of branching ratios for LFV processes considered in 
the present work. The muon conversion rate on the Ti atom is normalized 
on the muon capture rate on the Ti atom. For the muonium conversion, see 
Sec.~III F.}
\label{table1}
\vspace{.3in}
\begin{tabular}{ccc}
Mode  &  Branching ratio & Ref. 
\\     \tableline
$\mu \rightarrow e \gamma$ & $3.8 \times 10^{-11}$  &  \protect\cite{klaus}
\\ 
$\mu\rightarrow 3 e$ & $1.0 \times 10^{-12}$ & \protect\cite{klaus}
\\ 
$\mu^-$ + Ti $\rightarrow e^-$ + Ti & $6.1 \times 10^{-13}$  &
\protect\cite{klaus}
\\     \tableline
muonium conversion & $G_{MM}^{--} < 3.0 \times 10^{-3} G_F $ &
\protect\cite{mac}
\\
         & $G_{MM}^{+-} < 2.1 \times 10^{-3} G_F$   &
\\      \tableline
$\tau  \rightarrow e \gamma$ & $2.7\times 10^{-6}$  & \protect\cite{tau1}
\\
$\tau  \rightarrow \mu \gamma$ & $ 3.0\times 10^{-6}$  & \protect\cite{tau1} 
\\
$\tau \rightarrow 3 e$ & $2.9\times 10^{-6}$  & \protect\cite{tau2}
\\
$\tau \rightarrow 3 \mu$ & $1.9 \times 10^{-6}$  &  \protect\cite{tau2}
\\
$\tau \rightarrow  \mu e^+ e^-$ & $1.7 \times 10^{-6}$  &  \protect\cite{tau2}
\\
$\tau \rightarrow  e \mu^+ \mu^-$ & $1.8\times 10^{-6}$ & \protect\cite{tau2}
\\
$\tau \rightarrow  e \pi^0$ & $3.7 \times 10^{-6}$ & \protect\cite{tau3}
\\
$\tau \rightarrow  \mu \pi^0$ & $4.0 \times 10^{-6}$  &  \protect\cite{tau3}
\\
$\tau \rightarrow  e \eta$ & $8.2 \times 10^{-6}$ & \protect\cite{tau3}
\\
$\tau \rightarrow  \mu \eta$ & $9.6 \times 10^{-6}$   &  \protect\cite{tau3}
\\
$\tau \rightarrow  e \rho^0$ & $2.0 \times 10^{-6}$   & \protect\cite{tau2}
\\
$\tau \rightarrow  \mu \rho^0$ & $6.3 \times 10^{-6}$ & \protect\cite{tau2}
\\
$\tau \rightarrow  e \phi$ & $6.9 \times 10^{-6}$  & \protect\cite{tau2}
\\
$\tau \rightarrow  \mu \phi$ & $7.0 \times 10^{-6}$   & \protect\cite{tau2}
\end{tabular}
\end{table}

\vspace{.3in}

\begin{table}
\caption{
Limits on $(\delta_{ij}^l )_{LL,RR,LR,RL}$ from 
$l_i \rightarrow l_j + \gamma$ for 
$m_{\tilde{l}} = 100$ GeV and for 
different values of $x$ 
assuming there is no fortuitous cancellations among various terms. 
}
\label{table2}
\vspace{.3in}
\begin{tabular}{cccc}
Process & Constrained $\delta$ & $x$ & Limits
\\     \tableline
$\mu\rightarrow e\gamma$&$|(\delta_{12}^l)_{LL,RR}|$ & $0.3$ & 
$4.0\times 10^{-3}$
\\
&& $0.9$ & $7.6\times 10^{-3}$
\\
&& $3.0$ & $1.8\times 10^{-2}$
\\ 
&$|(\delta_{12}^l)_{LR,RL}|$& $0.3$ & $2.0\times 10^{-6}$
\\ 
&& $0.9$ & $2.3\times 10^{-6}$
\\ 
&& $3.0$ & $3.8\times 10^{-6}$
\\ \hline
$\tau\rightarrow e\gamma$&$|(\delta_{13}^l)_{LL,RR}|$& $0.3$ &
$2.5$
\\
&& $0.9$ & $4.7$
\\
&& $3.0$ & $11$
\\ 
&$|(\delta_{13}^l)_{LR,RL}|$& $0.3$ & $2.0\times 10^{-2}$
\\ 
&& $0.9$ & $2.4\times 10^{-2}$
\\ 
&& $3.0$ & $4.0\times 10^{-2}$
\\ \hline
$\tau\rightarrow \mu\gamma$&$|(\delta_{23}^l)_{LL,RR}|$&$0.3$ &
$2.4$
\\
&& $0.9$ & $4.5$
\\
&& $3.0$ & $10$
\\ 
&$|(\delta_{23}^l)_{LR,RL}|$& $0.3$ & $1.9\times 10^{-2}$
\\ 
&& $0.9$ & $2.3\times 10^{-2}$
\\ 
&& $3.0$ & $3.8\times 10^{-2}$
\end{tabular}
\end{table}
\vspace{.3in}

\begin{table}
\caption{
Upper limits for the branching ratios of $\tau\rightarrow l+P$ from the
constraints shown in Table II. }
\label{table3}
\vspace{.3in}
\begin{tabular}{ccc}
Process &  $x$  &  Braching Ratio \\ \hline
$\tau\rightarrow e\pi$ 
& $0.3$ & $0.61\times 10^{-12}$ \\
& $0.9$ & $0.38\times 10^{-9}$ \\
& $3.0$ & $0.24\times 10^{-8}$ \\ \hline
$\tau\rightarrow \mu\pi$ 
& $0.3$ & $0.56\times 10^{-12}$ \\
& $0.9$ & $0.34\times 10^{-9}$ \\
& $3.0$ & $0.20\times 10^{-8}$ \\ \hline
$\tau\rightarrow e\eta$ 
& $0.3$ & $0.20\times 10^{-12}$ \\
& $0.9$ & $0.13\times 10^{-9}$ \\
& $3.0$ & $0.81\times 10^{-9}$ \\ \hline
$\tau\rightarrow \mu\eta$ 
& $0.3$ & $0.19\times 10^{-12}$ \\
& $0.9$ & $0.12\times 10^{-9}$ \\
& $3.0$ & $0.67\times 10^{-9}$ \\ 
\end{tabular}
\end{table}
\vspace{.3in}

\begin{table}
\caption{
Upper limits for the branching ratios of $\tau\rightarrow l+V$ from the
constraints shown in Table II. }
\label{table4}
\vspace{.3in}
\begin{tabular}{ccc}
Process &  $x$  &  Braching Ratio \\ \hline
$\tau\rightarrow e\rho^0$ 
& $0.3$ & $0.32\times 10^{-7}$ \\
& $0.9$ & $0.42\times 10^{-7}$ \\
& $3.0$ & $0.42\times 10^{-7}$ \\ \hline
$\tau\rightarrow \mu\rho^0$ 
& $0.3$ & $0.30\times 10^{-7}$ \\
& $0.9$ & $0.38\times 10^{-7}$ \\
& $3.0$ & $0.34\times 10^{-7}$ \\ \hline
$\tau\rightarrow e\phi$ 
& $0.3$ & $0.29\times 10^{-8}$ \\
& $0.9$ & $0.39\times 10^{-8}$ \\
& $3.0$ & $0.43\times 10^{-8}$ \\ \hline
$\tau\rightarrow \mu\phi$ 
& $0.3$ & $0.26\times 10^{-8}$ \\
& $0.9$ & $0.35\times 10^{-8}$ \\
& $3.0$ & $0.36\times 10^{-8}$ \\ 
\end{tabular}
\end{table}
\vspace{.3in}

\begin{table}
\caption{Allowed ranges for the flavor conserving mass insertion 
$(\delta_{ii}^l)_{LR}$ from the anomalous magnetic moment of a lepton
for $m_{\tilde{l}} = 100$ GeV.
We show two cases $i=1 (e)$ and $i=2 (\mu)$ only, since the anomalous 
magnetic moment of a tau lepton is poorly measured.}
\label{table5}
\vspace{.3in}
\begin{tabular}{ccc}
$x$ & $(\delta_{11}^l)_{LR}$ & $(\delta_{22}^l)_{LR}$      
\\     \tableline
0.3 & $-3.0\times 10^{-2} \sim -2.3 \times 10^{-2}$ &
$-6.0\times 10^{-2} \sim 1.0 \times 10^{-2}$ 
\\
0.9 & $-3.6\times 10^{-2} \sim -2.7 \times 10^{-2}$ &
$-7.0\times 10^{-2} \sim 1.2 \times 10^{-2}$ 
\\
3.0 & $-5.9\times 10^{-2} \sim -4.5 \times 10^{-2}$ &
$-0.11 \sim 2.0 \times 10^{-2}$ 
\end{tabular}
\end{table}
\vspace{.3in}

\end{document}